\documentclass[]{article}
 \usepackage{epsfig}

\begin{document}

%%%%%%%%%%%%% style macros

\newcommand{\va}{\vspace{5mm}}
\newcommand{\vb}{\vspace{10mm}}
\newcommand{\vs}{\vspace{15mm}}

% \newcommand{\tc}[2]{\textcolor{#1}{#2}}       % 1 fg color, 2 text
% \newcommand{\cb}[2]{\colorbox{#1}{#2}}        % 1 bg color, 2 text
% \newcommand{\fb}[3]{\fcolorbox{#1}{#2}{#3}}   % 1 frame, 2 bg, 3 text

% \color{#1} changes fg color to named color #1
% \color[rgb]{#1,#2,#3} changes fg color to rgb color=(1,2,3).

% \parindent0mm
% \fboxsep3mm
% \fboxrule1mm

%%%%%%%  equation macros

\newcommand{\be}{\begin{equation}}
\newcommand{\ee}{\end{equation}}
\newcommand{\ba}{\begin{eqnarray}}
\newcommand{\ea}{\end{eqnarray}}
\newcommand{\NL}{\nonumber \\}

\newcommand{\binom}[2]{ \left( \begin{array}{c} #1 \\ #2 \end{array} \right)}
\newcommand{\E}[1]{ \langle #1 \rangle }
\newcommand{\ov}[1]{ \overline{#1} }

\newcommand{\QQ}{ \E{q} \E{\ov{q}} }

%%%%%%%%% FIG macro %%%%%%%%%%

\newcommand{\inplot}[4]{
 \begin{center}
 \leavevmode\epsfysize=#2mm \epsfbox{#1}
 \end{center}
 {#3} \quad {{\small #4}}
}

%%%%%%%  special macros

% \newcommand{\PG}{PineGreen}
% \newcommand{\PG}{OliveGreen}
% \newcommand{\cpg}{\color{\PG}}

%%%%%%%%% header  definition

% \def\fejlec{
%   \hrulefill \qquad {\bf T.S.Bir\'o: finiteness and statistics ...
%   \quad (\today) } \qquad  \hrulefill \ \
% }

%% normal pagestyle uses foilheadings

% \pagestyle{foilheadings}
%% \leftheader{{\color{RoyalPurple} \fejlec}}
%% \rightheader{{\color{RoyalPurple} \fejlec}}

%%%%%%%%%%%  front page

%%%%%%%%%%%%%%%%%%%%%%%% TITLE PAGE %%%%%%%%%%%%%%%%%%%

% \thispagestyle{empty}

% \draft

\title{Canonical enhancement as a result of Poisson distribution}

\author{T.S.~Bir\'o, \\ \quad \\ 
 MTA RMKI, H-1525 Budapest, P.O.Box 49  
}

\maketitle

\begin{abstract}
We point out that a certain finite size effect in heavy ion physics,
the canonical enhancement, is based on the difference of conservation
constrained pair statistics for Poisson and Gauss distributions,
respectively. Consequently it should occur in a wide range of phenomena,
whenever comparing rare and frequent events.

\end{abstract}

%\newpage

% \section{Introduction}

\va
It is often so, in particular in relativistic heavy ion collisions
observing thousands of newly produced particles in an event, that
only statistical information is available for tracing back characteristica
of an earlier stage of interacting matter. Aiming at the discovery
of quark gluon plasma it is especially important to be aware of
an unavoidable loss of information due to the very nature of statistics.
Selecting out, however, relevant control parameters even statistical
information can be used to make qualitative distinctions with respect to
the kind of matter existed for a while in such experiments, in particular
drawing conclusions about a possible equation of state.

\va
A truly interesting control parameter is the size of the reacting
system, which is often regarded as infinite in theory (as the so called
thermodynamical limit). In reality it is, however, 
finite and to a certain degree controllable
in experiments by varying the target nucleus and triggering measurements
by centrality (multiplicity, transverse energy) of a collision.
But even accepting the infinite size assumption (theoretically
in any high energy collision even infinitely many particle pairs can be
created for a while), there is a further distinction related to
the mean particle number in an event.

\va
Two famous, very characteristic limits can be considered: i) the limiting
case of frequent events producing many particles on the mean, 
approximately described by the Gauss distribution, 
and ii) very rare events with few (less than one)
particles per event as a mean value, generally covered by the Poisson
distribution. In this letter we point out that a characteristic
finite size effect in relativistic heavy ion collisions, named
{\em canonical enhancement}, can be understood as originating in the
difference between rare and frequent events, between sparse and
cupious particle production. Mathematically this reveals itself in
the difference between a large and a small mean particle number.
The large number case - as it is well
known - gives results equivalent to those stemming from a Gauss
distribution (central limit theorem of statistics).

\va
The latter equivalence is the basis of the equivalence of the canonical
and grand canonical approaches in the thermodynamical limit, while the
Poisson statistics with a small mean number plots the
difference between these approaches. The ratio of the canonical and
grand canonical result in small systems is less than one, the same
result can be obtained simply relying on the Poisson distribution
alone. The effect is in principle more general than the canonical
-- grand canonical distinction; processes and phenomena featuring
Poisson statistics occur also in physical situations far from
equilibrium. Conversely, observation of this deviation signals
truly a mesoscopic nature of the physical system (a fireball formed
in heavy ion collisions), but does not prove in itself equilibrium,
nor measures temperature or volume of the system. The mesoscopic
nature reflected by the 'canonical' suppression factor can be
extracted from mean particle numbers alone.

\va
The importance of this finite size effect for relativistic heavy ion
collisions, in particular its explanation power for finding more
strange particles per nucleon produced in ion - ion than in proton - proton
collisions (the so called strangeness enhancement), was first realized
by Redlich et. al.\cite{1,1a}. Originally presented as a constraint stemming
from the proper statistical treatment of a U(1) symmetry in the
canonical approach (whence the name canonical enhancement)\cite{3}, 
and referring to volume and temperature during the derivation of the
result, soon was it rederived on the basis of rate equations\cite{2},
and has been shown that there can be an underlying master equation
with the Poisson distribution as stationary solution.
Rafelski and Letessier emphasized that this phenomenon is strongly
related to pair statistics (to associated production of conserved
charge and anticharge) and it holds also out of chemical equilibrium
with general fugacity factors\cite{4}. Also a debate has been started
about the relevance of this effect on the strangeness enhancement
in heavy ion collisions, especially at CERN SPS using 40, 60 and 160
GeV/nucleon beam energies.

\va
Our aim in this letter is to rederive the canonical enhancement factor
relying on the Poisson distribution alone, not assuming even thermal
equilibrium in the background. One does not need to refer either to
temperature nor to volume in this derivation, -- even if also thermal
systems in a given volume may show Poisson statistics, this is not
their only possible origin. Then we apply this finite number suppression factor
to an analysis of particle production in the framework of the
sudden quark coalescence picture of hadron formation, ALCOR\cite{5}.
We recover qualitatively the linear coalescence assumption for
the low particle number case, i.e. that the number of composite
objects (mesons and baryons) are proportional to the product of the
numbers of its constituents (the quarks and antiquarks).

%%%%%%%%%%%%%%%%%%%%%%%%%%%% POISSON / GAUSS %%%%%%%%%%%%%%%%%%%%

% \section{Poisson vs. Gauss}

\va
From the probability $P(n)$ of getting $n$ particles of a certain
kind the expectation value $\langle n \rangle$ and higher moments
like quadratic spread can easily be obtained by the use of the
following generating function:

\be
 Z(\gamma) = \sum_n P(n) e^{\gamma n}.
\ee
The expected (mean) number becomes
\be
 \langle n \rangle = \left. \frac{\partial}{\partial \gamma} \log Z \right|_{\gamma=0}.
\ee
Under certain circumstances we are particularly interested in pairs of
particles carrying a sum or a difference of a given physical quantity,
than in the one-particle distribution. On the basis of independent
one-particle distributions the pair statistics and the distribution of the
sum can be calculated. Without any constraint on the difference, the
distribution of the sum is described by the convolution of the 
corresponding distributions:
\be
P_U(s) = \sum_{n,m} P_1(n)  P_2(m)  \delta_{m+n,s} =
\sum_n P_1(n) P_2(s-n).
\ee
This is, however, correct only if there is no information on the
difference. In particular the production of particles and
antiparticles happens always in pairs due to charge (and further)
conservation laws, and as a consequence, even if the production
process is statistical, their difference is bounded to be zero.
In this case we obtain another distribution of the sum:
\be
P_C(s|0) = \frac{1}{P_0} \sum_{m,n} P_1(n) P_2(m) \delta_{m+n,s} \delta_{m-n,0}
= \frac{P_1(s/2)P_2(s/2)}{\sum_n P_1(n)P_2(n)}.
\ee
This is a conditional probability with the factor $P_0$ being the probability of
getting zero difference in the number statistically, which ensures the
normalization of the result to one.
Fig.1 shows a geometrical interpretation of the distinction between these two
pair statistics: the unconstrained distribution can be obtained by adding
all points in $(n,m)$ space over a given bin of the diagonal representing
a given sum $s=n+m$, while in case of the fixed zero difference only hits
in the diagonal stripe counts (and the ratio of the stripe to the total
area reflects the $1/P_0$ normalization factor).

%%%%%%%%%%%%%%%%%%%%%%%% FIG. 1 %%%%%%%%%%%%%%%%%%%%%%%%%%%
\newpage
\vspace*{20mm}

 \label{FIG1}
 \inplot{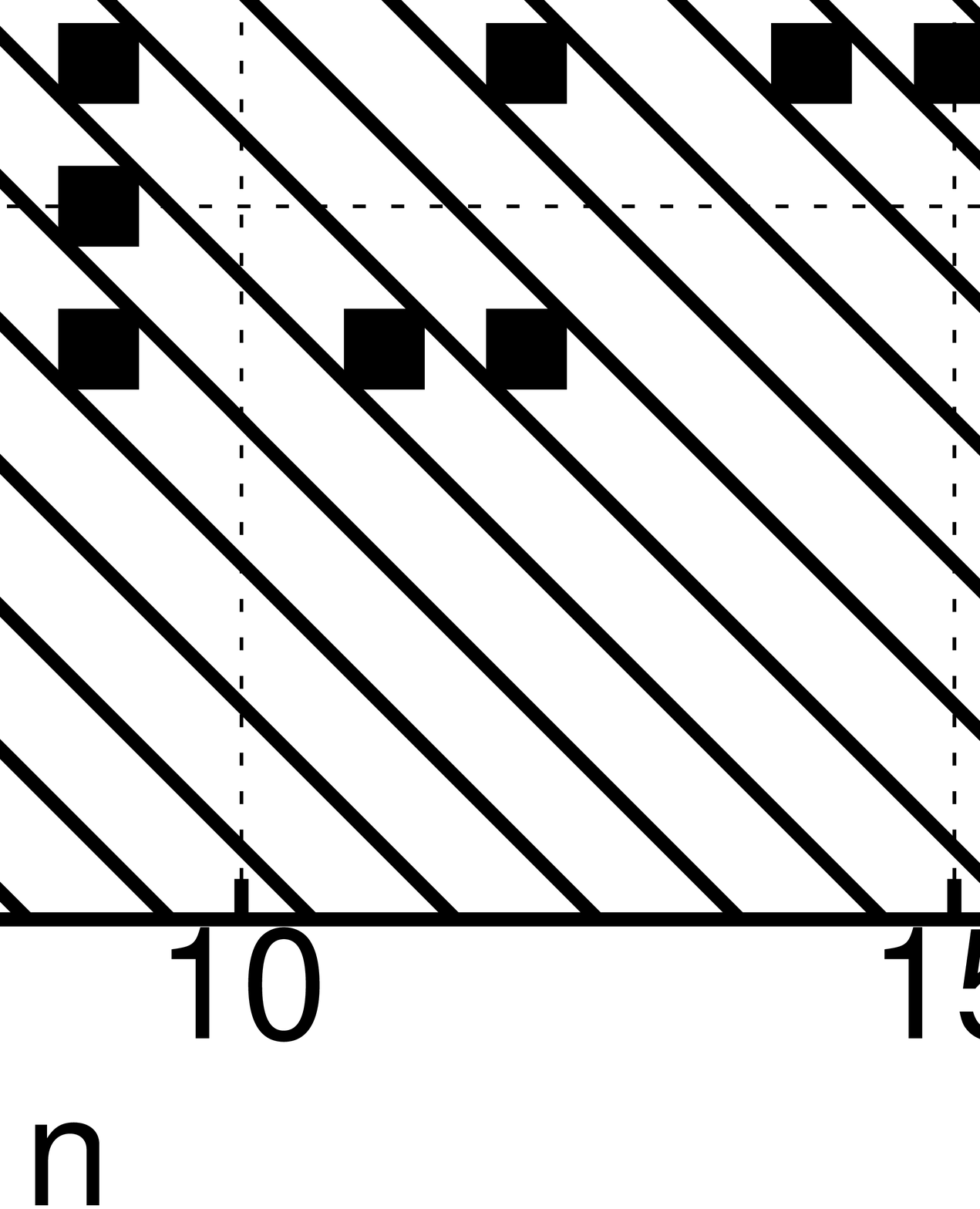}{25}{Fig.1}{
  Demonstrating the constrained (main diagonal stripe) 
  and uncostrained (the whole square) distribution of
  the sum of two Poisson deviates with mean value 10 for each.  
 }
%%%%%%%%%%%%%%%%%%%%%%%%%%%%%%%%%%%%%%%%%%%%%%%%%%%%%%%%%%%%%

\va
Applying this idea to restless coalescence of quarks into hadrons, we still
constrain the difference to zero, but this time due to a confinement
principle. The number of coalesced hadrons is the half sum in this case.
Caring for pions only in a simplified world of light quarks and antiquarks,
we arrive at having $n$ from both, with zero difference and a sum of $2n$
forming exactly $n$ pions. The generating function of the constrained
distribution $P_C(s|0)$ in this case reads as
\be
Z_C(\gamma) = \frac{1}{P_0} \sum_n P_q(n) P_{\overline{q}}(n) e^{2\gamma n}.
\ee
Assuming Poisson distribution for the quarks this generating function
becomes
\be
Z_C(\gamma) =  \frac{1}{P_0} \sum_n \frac{\E{q}^n}{n!}e^{-\E{q}}
\frac{\E{\ov{q}}^n}{n!}e^{-\E{\ov{q}}} e^{\gamma n} = 
\frac{1}{P_0} e^{-\E{q} - \E{\ov{q}}} \, I_0(2e^{\gamma}\sqrt{\QQ}).
\ee
with $I_0(x)$ being the Bessel function of the first kind with imaginary argument.
The normalization factor $1/P_0$ can be obtained from $Z(0)=1$, giving
\be
Z_C(\gamma) = \frac{I_0(2e^{\gamma}\sqrt{\QQ})}{I_0(2\sqrt{\QQ})}.
\ee
The expectation value of the pions is
\be
\E{ \pi }_P = \frac{1}{2} \E{s} = 
\sqrt{\QQ} \quad \frac{I_1(2\sqrt{\QQ})}{I_0(2\sqrt{\QQ})},
\ee
with the label $P$ reminding us to the Poisson distribution.
For small mean number of quarks this leads to the product assumption
of the simple linear coalescence model:
\be
\E{\pi}_P \approx 2 \QQ.
\ee
For large mean numbers on the other hand the ratio of the Bessel functions,
$I_1/I_0$ approaches one. This latter factor is the canonical suppression
factor.

\va
As opposed to the above analysis of the Poisson distribution, for the
Gauss distribution both the unconstrained and the zero difference
constrained pair statistics leads again to a Gauss distribution
(but with a larger width). Assuming Gauss distributed quark and
antiquark numbers with respective square widths equal to the mean
values, as it is typical for equilibrium ideal gases, the expectation
value of the half sum, the pion number becomes the harmonic mean of
the expectations (because by convolution of Gauss distributions the
inverse square widths are additive):
\be
\E{\pi}_G = \frac{ 2\E{q} \E{\ov{q}} }{ \E{q} + \E{\ov{q}} }.
\ee
In case of zero baryon charge $\E{q} = \E{\ov{q}}$ and these formuli
reduce to a very simple result:
\be
\E{\pi}_G = \E{q}; \qquad \E{\pi}_P = \E{q} \quad
 \frac{ I_1(2\E{q}) }{ I_0(2\E{q}) }.
\ee
As a consequence one can express canonical enhancement in this case
as a relation between Poisson and Gauss expectation values.

\va
It is particularily interesting when we compare rare particles,
for example $K^+$ mesons coalesced from Poisson distributed 
$u$ and $\ov{s}$ quarks with copiusly produced $\pi^+$ pions
glued from $u$ and $\ov{d}$ quarks. The expected meson numbers,
\be
\E{K^+}_P = \sqrt{u\ov{s}} \quad \frac{I_1(2\sqrt{u\ov{s}})}{I_0(2\sqrt{u\ov{s}})},
\ee
for kaons and
\be
\E{\pi^+}_G = \frac{2 u \ov{d}}{u+\ov{d}}
\ee
for pions (with $u$, $\ov{s}$ and $\ov{d}$ denoting here the expectation
values of the respective quark numbers) for a baryon free and strangeness
free, and isotopically symmetric fireball due to 
$\ov{d} = d = \ov{u} = u$ and $\ov{s} = s = f \ov{d}$ with a fixed
strangeness ratio $f = \ov{d}/\ov{s}$ reduce to
\be
\E{K^+}_P = u \sqrt{f} \quad \frac{I_1(2u\sqrt{f})}{I_0(2u\sqrt{f})}
\ee 
and
\be
\E{\pi^+}_G = u.
\ee
In this simplified scenario the kaon to pion ratio  can be expressed as
a function of the pion number and so can be compared with experimental
results. Of course the result can only be qualitative on this level, since
by using the above assumptions for the quark and antiquark numbers,
one is bound to predict the same ratio for $K^+/\pi^+$ and for $K^-/\pi^-$. 
It is not quite fulfilled in experiments: comparing the heavy ion results 
with those of proton - proton collisions, these ratios increase by a factor of
roughly $2$ and $1.6$ respectively\cite{ADD}.

\va
The result of this simple idea can be inspected in Fig.2, where
the normalized ratio $\E{K^+}_P/(\E{\pi^+}_G\sqrt{f}$ is plotted
versus the scaled pion number, $2\E{\pi^+}_G\sqrt{f}$. This is
exactly the canonical suppression factor. Using the value $f \approx 0.5$
for estimating the Wroblinski factor $f$ one concludes
that the kaon / pion ratio falls to its half at an expected pion
number of about $0.5$ as compared to large pion numbers. This value corresponds
to a situation before resonance decays, so it is not directly
observable in experiments, but correcting for resonance decays is
always possible in the framework of a tehoretical model. We used
ALCOR for this purpose and concluded that the kaon-pion ratios normalized
to the $pp$ case do not change much due to resonance decay, as the
unnormalized ratios do.

\vspace{40mm}
\label{FIG2}
\inplot{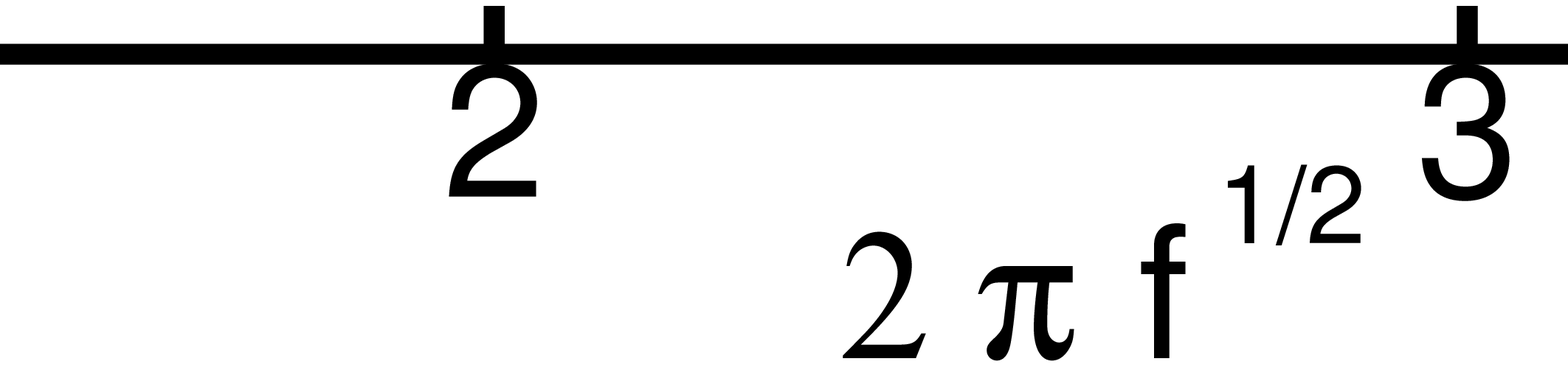}{35}{Fig.2}{
  Canonical suppression factor obtained from the expectation value
  of the sum of two (numerically simulated) Poisson distributions. 
  It coincides with the analytically derived ratio of two Bessel functions $I_1/I_0$.
}

\va
Assuming that both pions and kaons are constructed from Poisson
distributed quarks, their ratio changes between two limiting
values corresponding to few and many quarks respectively:
\be
\frac{\E{K}_P}{\E{\pi}_P} = \sqrt{f} \quad
\frac{I_1(2u\sqrt{f})}{I_0(2u\sqrt{f})} \quad \frac{I_0(2u)}{I_1(2u)},
\ee
gives $f$ for $u \ll 1$ and $\sqrt{f}$ for $u \gg 1$. Considering
the light quark (pion) number before resonance decay the former
case is realized in $pp$ collisions (according to ALCOR $\E{\pi} \sim 0.1$)
and the latter case in $PbPb$ collisions ($\E{\pi} \sim 10$, before resonance
decay).  The double ratio,
\be
\frac{(K/\pi)_{PbPb}}{(K/\pi)_{pp}} \approx \frac{1}{\sqrt{f}}
\ee
would be about $1.4$, which is to be compared with the experimental
values $1.6$ for negative and $2$ for positive kaon to pion ratios.
The transition is around $\E{\pi} \approx 0.7 \ldots 1.4$, probably
occuring in a collision of relatively light ions.

% \section{Conclusion}

\va
In conclusion we pointed out that the canonical suppression factor
of a ratio of Bessel functions can be derived solely relying on properties
of the Poisson distribution, and hence is only as much related to
thermal equilibrium properties as the latter realizes a Poisson
distribution for rare particles. This distribution, however, may stem
also from a series of dynamical events and therefore the corresponding
factor should show in a wide class of phenomena. We gave a rough
estimate of reflecting this size (expected number) dependence
in heavy ion collisions by the strangeness enhancement in the framework
of a quark coalescence hadronization model, ALCOR.

\va
{\bf Acknowledgement}
\quad Discussions with K. Redlich, J. Knoll, P.L\'evai and J. Zim\'anyi
are gratefully acknowledged. This work has been supported by the
Hungarian National Science Fund OTKA (T034269).

\end{document}